\documentclass[aps,prd,preprintnumbers,nofootinbib,twocolumn]{revtex4-2}
\usepackage{bm}
\usepackage{enumitem}   
\usepackage{latexsym}
\usepackage[dvipsnames]{xcolor}
\usepackage{dcolumn}
\usepackage{amsmath,amsfonts,amssymb}
\usepackage{graphicx,float}
\usepackage{appendix}
\usepackage[hyperfootnotes=true]{hyperref}
\usepackage{mciteplus}
\usepackage{subcaption}
\usepackage{caption}
\usepackage{bbold}
\usepackage{slashed}
\usepackage{mathbbol}
\DeclareSymbolFontAlphabet{\amsmathbb}{AMSb}%
\usepackage{amsthm}
\usepackage{hyperref}
\usepackage{color}
\usepackage{mathrsfs}
\usepackage{setspace}
\usepackage{orcidlink}
\usepackage{aas_macros}

\DeclareMathAlphabet{\mathpzc}{OT1}{pzc}{m}{it}
\usepackage{calligra}
\DeclareMathAlphabet{\mathcalligra}{T1}{calligra}{m}{n}
\DeclareFontShape{T1}{calligra}{m}{n}{<->s*[2.2]callig15}{}

\def\be {\begin{equation}}
\def\ee {\end{equation}}
\def\bea {\begin{eqnarray}}
\def\eea {\end{eqnarray}}
\def\bc {\begin{center}}
\def\ec {\end{center}}
\def\bfg {\begin{figure}}
\def\efg {\end{figure}}
\def\bi {\begin{itemize}}
\def\ei {\end{itemize}}

\def\beq{\begin{equation}}
\def\eeq{\end{equation}}
\def\br{\begin{eqnarray}}
\def\er{\end{eqnarray}}
\newcommand{\eel}[1] {\label{#1}\end{equation}}

\newcommand{\bdm}{\begin{displaymath}}
\newcommand{\edm}{\end{displaymath}}

\begin{document}

\title{ A covariant tapestry of linear GUP, metric-affine gravity, their Poincar\'e algebra and entropy bound}


\author{Ahmed Farag Ali $^\nabla$$^{\triangle}$\orcidlink{0000-0001-8944-6356}}
\email[email: ]{aali29@essex.edu}

\author{Aneta Wojnar $^\odot$ \orcidlink{0000-0002-1545-1483}}
\email[E-mail: ]{awojnar@ucm.es}

\affiliation{{$^{\nabla}$Essex County College, 303 University Ave, Newark, NJ 07102, United States }}
\affiliation{ $^{\triangle}$ Dept. of Physics, Benha University, Benha 13518, Egypt}

\affiliation{$^\odot$Department of Theoretical Physics \& IPARCOS, Complutense University of Madrid, E-28040, 
Madrid, Spain}

\begin{abstract}
\par\noindent
Motivated by the potential connection between metric-affine gravity and  linear Generalized Uncertainty Principle (GUP) in the phase space, we develop a covariant form of linear GUP and an associated modified Poincaré algebra, which exhibits distinctive behavior, nearing nullity at the minimal length scale proposed by linear GUP. We use  3-torus geometry to visually represent linear GUP within a covariant framework. The 3-torus area provides an exact geometric representation of Bekenstein's universal bound. We depart from  Bousso's approach, which adapts  Bekenstein's bound by substituting the Schwarzschild radius ($r_s$) with the radius ($R$) of the smallest sphere enclosing the physical system, thereby basing the covariant entropy bound on the sphere's area. Instead, our revised covariant entropy bound is described by the area of a 3-torus, determined by both the inner radius $r_s$ and outer radius $R$ where $r_s\leq R $ due to gravitational stability. This approach results in a more precise geometric representation of Bekenstein's bound, notably for larger systems where Bousso's bound is typically much larger than Bekensetin's universal bound. Furthermore, we derive an equation that turns the standard uncertainty inequality into an equation when considering the contribution of the  3-torus covariant entropy bound, suggesting a new avenue of quantum gravity.


\end{abstract}

\maketitle

\section{introduction}
\par\noindent   
Several approaches to quantum gravity, including string theory, black hole physics, and quantum geometry, suggest the existence of a minimum measurable length. This concept, critical in theories like String Theory and Black Hole physics, poses significant questions about space-time's fundamental structure. In String Theory, for instance, strings are theorized to not interact at distances smaller than their inherent size, suggesting a minimal measurable length. This assumption leads to the development of the Generalized Uncertainty Principle (GUP) \cite{Amati:1988tn,Scardigli:1999jh,snyder1947quantized,Adler:2001vs,maggiore1993generalized,Capozziello:1999wx,Kempf:1994su}. 

\noindent
The linear GUP, introduced in \cite{Ali:2009zq}, represents a notable advancement in line with Doubly Special Relativity (DSR) \cite{Magueijo:2001cr,Cortes:2004qn}. Applicable in both Euclidean and Pseudo-Euclidean spacetimes, the linear GUP has led to insights into the quantum/discrete nature of space in non-relativistic \cite{Ali:2009zq}, relativistic \cite{Das:2010zf}, and curved spacetime contexts \cite{Das:2020ujn}. Consistent with DSR principles, linear GUP has been instrumental in introducing the quantization of space, a complement to the traditional quantization of energy. This has significantly impacted the solutions of wavefunction in quantum mechanics, allowing for the quantization of both spatial and energy parameters through the linear GUP framework in fundamental quantum equations like the Schrödinger, Klein-Gordon, and Dirac equations \cite{Ali:2009zq,Das:2010zf, Ali:2011fa}. This dual quantization approach marks a substantial advancement in understanding the quantum structure of space-time, bridging quantum mechanics and gravitational theories.

\noindent
The applications of GUP span a broad spectrum of fields and energy scales, ranging from atomic systems \cite{Ali:2011fa, Das:2008kaa}, quantum optics \cite{Pikovski:2011zk,Kumar:2017cka}, and cold atoms \cite{Tino:2020dsl,Gao:2016fmk}, to superconductivity and the quantum Hall effect \cite{Das:2011tq}, gravitational bar detectors \cite{Marin:2013pga}, gravitational waves \cite{Bosso:2018ckz,Feng:2016tyt,Moussa:2021gxb,Moussa:2021qlz}, and even extending to gravitational decoherence \cite{Petruzziello:2020wkd}, composite particles \cite{Kumar:2019bnd}, astrophysical systems \cite{Moradpour:2019wpj,Vagnozzi:2022moj}, Baryon asymmetry \cite{Das:2021nbq}, gravitational tests \cite{Hemeda:2022dnd}, and condensed matter systems \cite{Iorio:2022ave,Iorio:2017vtw}. Different theoretical approaches support linear GUP such as considering background coordinate time as a quantum degree of freedom coupled with mass-energy \cite{Singh:2023nvu},  discrete spacetime model of a graphene analog system \cite{Shah:2023xgj}.

The study \cite{Majumder:2011ad} demonstrates that linear GUP exhibits non-unitary properties, highlighting the significance of non-unitary processes in quantum mechanics. Non-unitary mechanics are essential for explaining various phenomena, including quantum decoherence, which describes the evolution of quantum states into classical states due to environmental interactions \cite{Zurek:2003zz}; the dynamics of open quantum systems influenced by their surroundings \cite{breuer2002theory}; the development of post-quantum theories integrating quantum mechanics with gravity \cite{Penrose:1996cv}; the examination of PT-symmetric quantum mechanics, which studies non-Hermitian systems yielding real eigenvalues under parity-time symmetry \cite{Bender:2007nj}; and the exploration of quantum measurement and wave function collapse through models like GRW, which propose spontaneous localization to achieve measurement outcomes without observer intervention \cite{Ghirardi:1985mt}. Furthermore, these non-unitary aspects contribute to the understanding of the information loss paradox in black holes, challenging the principle of unitary evolution and promoting research in quantum gravity \cite{Hawking:1976ra}. \\

In comparing linear and quadratic GUP models, \cite{Giardino:2020myz} found quadratic GUP more consistent with full cosmological data, while linear GUP better explained late cosmological phenomena. Contrarily, \cite{Das:2021nbq} argued linear GUP could account for observed baryon asymmetry without necessitating a negative GUP parameter, unlike its quadratic counterpart that implies an unphysical imaginary minimal length. Essentially, while quadratic GUP necessitates a negative parameter in order to explain the Baryon asymmetry, leading to non-physical implications for the minimal measurable length, linear GUP supports a baryon asymmetry explanation with a physical, real minimal length. Baryon asymmetry, pivotal for cosmological models from Big Bang Nucleosynthesis to cosmic structure, underscores potential new physics beyond the Standard Model, including mechanisms like CP violation \cite{Dine:2003ax, Canetti:2012zc, Riotto:1999yt}. This could indicate the requirement to integrate both linear and quadratic versions of the GUP to thoroughly interpret full cosmological data and Baryon asymmetry, based on a real minimal measurable length.

GUP models are significant as they introduce modifications in state equations and microscopic variables \cite{Brau:1999uv,moussa2015effect,rashidi2016generalized,belfaqih2021white,mathew2021existence,hamil2021new,gregoris2022chadrasekhar,barca2022comparison,Nouicer:2007jg,Pedram:2011gw,Barman:2023hyb,Shababi:2020evc,Shababi:2020ycd}. These models predict potential quantum gravity effects observable at various scales, with theories often approximating a minimum length scale akin to the Planck length ($\ell_{Pl}= \sqrt{\frac{\hbar G}{c^3}}$), emphasizing quantum gravity's role in the cosmic framework \cite{Pachol:2023bkv,Pachol:2023tqa,Kozak:2023vlj,chang2002exact,chang2002effect,segreto2023extended,Easther:2001fi}. For a more comprehensive understanding of the GUP, its varied phenomenology, and experimental implications, there are detailed reviews available \cite{Addazi:2021xuf,Hossenfelder:2012jw,Bosso:2023aht,Nouicer:2013poa,Scardigli:2022qes}.

\noindent
Subsequent research has uncovered a compelling relationship between linear GUP and metric-affine gravity \cite{Wojnar:2023bvv}. This connection deepens our understanding of the interplay between quantum mechanics and gravity, indicating necessary modifications in state equations and microscopic variables within Palatini gravity \cite{Wojnar:2024xdy}. Despite General Relativity's (GR) success in explaining many phenomena, its limitations in dark matter \cite{rubin1980rotational}, dark energy \cite{huterer1999prospects}, and early cosmological inflation \cite{copeland2006dynamics,nojiri2007introduction,nojiri2017modified,nojiri2011unified,capozziello2008extended,CANTATA:2021ktz} necessitate the exploration of modified gravity theories. This exploration not only addresses these limitations but also enriches our comprehension of the universe's fundamental nature.

\noindent
In section \ref{Sec:GUP-Modifiedgravity}, we explore in depth the connection between the linear GUP and Palatini-like gravity theories. In section \ref{sec:GUP-Poincare}, our focus is on developing a covariant form of linear GUP along with a modified Poincaré algebra, emphasizing the unique properties of this algebra at the minimal length scale indicated by linear GUP. We specifically examine the modified Poincaré algebra, noting its tendency to approach nullity at the minimal measurable scale. An intriguing aspect of this modification is the exception found in the commutation relation between position and Lorentz generators, which may hint at fundamental fixed-background algebra characteristics inherent to quantum gravity. Acknowledging the recently evolved correspondence between the Bekenstein entropy bound and the linear GUP, we provide in section \ref{sec:GUP-Bekenstein} a visual illustration of linear GUP. This representation is achieved through the precise geometry of a 3-torus, offering an exact geometric representation of Bekenstein's universal bound, particularly beneficial for larger systems. Additionally, we derive an equation transforming the standard uncertainty inequality into an equation, suggesting a new perspective on quantum gravity. Conclusions are discussed in the section \ref{sec:conclusions}.

\section{Correspondence between Ricci-based gravity and linear GUP} \label{Sec:GUP-Modifiedgravity}
\noindent
We will reassess the link between modified gravity and the linear GUP. To begin with, we will explore fundamental concepts related to Ricci-based gravity. Following that, we will reconsider the established correlation, investigating the deformation of the phase space in Palatini-like proposals and examining its consequences for thermodynamics. Consider a category of metric-affine gravity models referred to as Ricci-based gravity theories (cf. \cite{alfonso2017trivial}). The associated action for this class is given by:
\begin{equation} \label{eq:actionRBG}
\mathcal{S}=\int d^4 x \sqrt{-g} \mathcal{L}_G(g_{\mu\nu},\mathcal R_{\mu\nu}) +
\mathcal{S}_m(g_{\mu\nu},\psi_m) .
\end{equation}
Here, $g$ represents the determinant of the space-time metric $g_{\mu \nu}$, and $\mathcal R_{\mu \nu}$ denotes the symmetric Ricci tensor. Notably, the latter is independent of the metric, constructed solely with the affine connection $\Gamma \equiv \Gamma_{\mu\nu}^{\lambda}$. We introduce the object ${M^ \mu}_{\nu} \equiv g^{\mu \alpha}\mathcal R_{\alpha\nu}$ to formulate the gravitational Lagrangian $\mathcal{L}_G$, constructed as a scalar function using powers of traces of ${M^ \mu}_{\nu}$. On a different note, the matter action is given by:
\begin{equation}
\mathcal{S}_m=\int d^4 x \sqrt{-g} \mathcal{L}_m(g_{\mu\nu},\psi_m).
\end{equation}
In this formulation, the matter action is minimally coupled to the metric, ignoring the antisymmetric part of the connection (torsion), similar to the treatment of minimally coupled bosonic fields. This simplification extends to fermionic particles, such as degenerate matter, effectively described by a fluid approach, exemplified by the perfect fluid energy-momentum tensor \cite{alfonso2017trivial}. Similarly, focusing on the symmetric part of the Ricci tensor avoids potential ghostlike instabilities \cite{Borowiec:1996kg,Allemandi:2004wn,beltran2019ghosts,jimenez2020instabilities}. This approach accommodates various gravity theories, including GR, Palatini $f(\mathcal R)$ gravity, Eddington-inspired Born-Infeld (EiBI) gravity \cite{vollick2004palatini}, and its numerous extensions \cite{jimenez2018born}. The gravitational action encompasses theories that, despite possessing intricate field equations, can be conveniently reformulated, as demonstrated in \cite{jimenez2018born}:
\begin{equation} \label{eq:feRBG}
{G^\mu}_{\nu}(q)=\frac{\kappa^2}{\vert \hat{\Omega} \vert^{1/2}} \left({T^\mu}_{\nu}-\delta^\mu_\nu \left(\mathcal{L}_G + \frac{T}{2} \right) \right) .
\end{equation}
Here, $\vert\hat{\Omega}\vert$ denotes the determinant of the deformation matrix, and $T$ is the trace of the energy-momentum tensor of matter fields. The Einstein tensor ${G^\mu}_{\nu}(q)$ is associated with a tensor $q_{\mu\nu}$, where the connection $\Gamma$ assumes the Levi-Civita connection of $q_{\mu\nu}$:
\begin{equation}
\nabla_{\mu}^{\Gamma}(\sqrt{-q} q^{\alpha \beta})=0.
\end{equation}
For this formalism, the tensor $q_{\mu\nu}$ is related to the space-time metric $g_{\mu\nu}$ through:
\begin{equation}\label{eq:defmat}
q_{\mu\nu}=g_{\mu\alpha}{\Omega^\alpha}_{\nu}.
\end{equation}
The deformation matrix ${\Omega^\alpha}_{\nu}$ is theory-dependent, determined by the gravitational Lagrangian $\mathcal{L}_G$. Importantly, these theories yield second-order field equations, reducing to GR counterparts in vacuum (${T_\mu}^{\nu}=0$). This implies no extra degrees of freedom propagate in these theories beyond the usual two polarizations of the gravitational field. Of particular interest is the nonrelativistic limit of the field Eqs. \eqref{eq:feRBG}. In Palatini $f(\mathcal R)$ \cite{Toniato:2019rrd} and EiBI \cite{banados2010eddington,pani2011compact} gravities, the Poisson equation takes the form:
\begin{equation}\label{poisson}
\nabla^2\phi = \frac{\bar\kappa}{2}\Big(\rho+\sigma\nabla^2\rho\Big).
\end{equation}
Here, $\phi$ is the gravitational potential, $\bar\kappa=8\pi G$, and $\sigma$ is a theory parameter. The expressions for $\sigma$ are $\sigma=2\beta$ for Palatini $f(\mathcal R)$, with $\beta$ accompanying the quadratic term, and $\sigma=\epsilon/2$ for EiBI, where $\epsilon=1/M_{BI}$ and $M_{BI}$ is the Born-Infeld mass. It is noteworthy that the similarity in the Poisson equation between these two gravity proposals is not coincidental; the EiBI gravity in the first-order approximation reduces to Palatini gravity with the quadratic term \cite{pani2012surface}. Furthermore, only the quadratic term $\mathcal R^2$ influences the non-relativistic equations, as higher curvature scalar terms enter the equations at the sixth order \cite{Toniato:2019rrd}. In \cite{Wojnar:2023bvv} it was demonstrated that one can consider the Palatini-like theories with the non-relativistic limit given by \eqref{poisson} as the linear GUP. That is, the additional term present in the Poisson equation \eqref{poisson} can be interpreted as a modification to the Fermi gas at finite temperature. This modification arises when considering a deformation of the phase space given by the integral
\begin{equation}\label{sumint}
\frac{1}{(2\pi \hbar)^3} \int \frac{d^3xd^3p}{(1-\gamma p)^{d}},
\end{equation}
where the specific case of $d=1$ corresponds to the non-relativistic limit of the Palatini-like theories of gravity and the deformation parameter is related to the quadratic Palatini $f(\mathcal R)$ gravity by the expression
\begin{equation}
\gamma = \frac{4\pi G}{K_2}\beta\,\,\,\,\text{and}\;\;\; K_2 = \frac{3}{\pi} \frac{h^3N_A^2}{m_e \mu_e^2}.
\end{equation}
Similar to the GUP featuring linear $p$-modifications \cite{cortes2020deformed,Ali:2009zq,Ali:2011fa,Ali:2011ap,abac2021modified,vagenas2019gup}, our approach incorporates a deformed phase space measure characterized by the deformation parameter $\gamma$. In the context of GUP, this deformation is determined through the application of the Liouville theorem \cite{Ali:2011ap}. The deformation of the phase space \eqref{sumint} is a consequence of the fact that the unit volume of the space has to be invariant upon the change in the momentum for every state. In other words, laws of physics must be the same, independently of any change in space and time. Therefore, one expects that such a change in the volume element must arise from a generalized commutation relation between momenta and positions. It turns out that the deformation \eqref{sumint} introduced by metric-affine gravity is a special case of linear GUP  studied in \cite{Ali:2011ap}:
\begin{equation}
    \frac{V}{(2\pi \hbar)^D} \int \frac{d^Dxd^Dp}{(1-\alpha p)^{D+1}}.
\end{equation}
To see that, notice that up to the linear order approximation in the parameter $\gamma$ and $\alpha$ we will deal with the same modifications of the thermodynamic quantities\footnote{Consider
\begin{equation*}
    (1-\alpha p)^{-(D+1)}\approx 1+(D+1)\alpha p + \mathcal{O}^2(\alpha)
\end{equation*}
and 
\begin{equation*}
    (1-\gamma p)^{-1}\approx 1+\gamma p + \mathcal{O}^2(\gamma)
\end{equation*}
which appear in the partition function and thermodynamic quantities derived from it \cite{Wojnar:2023bvv}.
} if we consider the rescalling as $\gamma = (D+1)\alpha$. Consequently, the effective $\hbar$ depends on the momentum $p$ in the generalized uncertainty relation, resulting in a momentum-dependent size of the unit cell for each quantum state in phase space. Therefore, the linear corrections in \eqref{sumint} arising from minimal length are related to the quadratic correction to the gravitational Lagrangian of a metric-affine gravity, that is,
\begin{equation}\label{quadratic}
    f(\mathcal R)= \mathcal R + \beta \mathcal R^2 + \mathcal{O}^3(\mathcal R),
\end{equation}
where the curvature scalar is built of both, independent connection $\Gamma$ and metric $g$
\begin{equation}
    \mathcal R=g^{\mu\nu}\mathcal R_{\mu\nu}(\Gamma) .
\end{equation}
Therefore, considering the $f(\mathcal R)$ gravity as the simplest example of the Ricci-based family to get insight into our further analysis, let us briefly discuss properties resulting from this model. The Palatini curvature scalar turns out to depend on the trace of the energy-momentum tensor $T$, that is,
\begin{equation}
    \mathcal R = - \kappa^2 T,
\end{equation}
for the quadratic model \eqref{quadratic}. It results from the trace of the field equations (obtained by taking the variation with respect to the metric $g$)
\begin{equation} 
f'(\mathcal {R})\mathcal{R}_{\mu\nu}-\frac{1}{2}f(\mathcal{R})g_{\mu\nu}=\frac{8\pi G}{c^4} T_{\mu\nu}.
\end{equation}
Considering a system, whose matter is described by the perfect fluid, with $p<<\rho/c^2$, we deal with $\mathcal R\sim \rho$. Recent investigations have revealed that standard quantum mechanics might be derived from the full mathematical structure of the Lorentz transformation, which notably includes its superluminal component. This approach leads to the emergence of non-deterministic dynamics, along with the introduction of complex probability amplitudes and diverse trajectories \cite{Dragan:2019grn}. Considering the potential correlation between modified gravity and the linear GUP in the phase space  \cite{Wojnar:2023bvv}, it is conceivable that both these theories could be expressed using a modified Poincaré's algebra. This concept will be further examined in the subsequent section of our study.

\section{Covariant form of linear GUP and its symmetry} \label{sec:GUP-Poincare}
\noindent
The correspondence between linear GUP and modified gravitational theories motivates the development of a covariant form of the linear GUP, applicable across any spacetime metric ($g_{\mu\nu}$). The covariant form of linear GUP which may correspond to the Palatini-like gravity could take the following form:
\begin{equation}
  [x_{\mu}, p_{\nu}] = i \hbar \left[ g_{\mu \nu} - \alpha \left( p g_{\mu \nu} + \frac{p_{\mu} p_{\nu}}{p} \right) \right],
\end{equation}
where $g_{\mu \nu}$ is a four-dimensional spacetime metric and $\alpha= \alpha_0 \ell_{Pl}/\hbar$ such that $\alpha_0$ is a dimensionless parameter that quantifies the minimal measurable length. This  commutator satisfies the Jacobi identity and ensure that  $[x_{\mu},x_{\nu}]=0=[p_{\mu},p_{\nu}]$. Defining 
\begin{eqnarray}
x_{\mu} &=& x_{0\mu} ,  \label{position}\\
p_{\mu} &=& {p_{0\mu}}(1 - \alpha p_0), \label{momentum}
\end{eqnarray}
where  $x_{0\mu}$  and  $p_{0\nu}$ satisfy the standard commutation relations: $[x_{0\mu}, p_{0\nu}] = i g_{\mu\nu} \hbar$, and $p_0^2=p_{0\mu}p_0^{\mu}$. In this context,  $p_{0\mu}$ represents the low-energy momentum  expressed in position space as  $p_{0\mu} = -i \frac{\partial}{\partial x_{0\mu}}$), whereas  $p_{\mu}$ denotes the high-energy momentum, with  $p_0$ being the norm of the  $p_{\mu}$  vector. In the work of Todorinov et al. \cite{Todorinov:2018arx}, the Poincaré algebra was modified for the quadratic Generalized Uncertainty Principle (GUP), whereas in our research, we aim to develop the Poincaré algebra for the linear GUP. Using equations (\ref{position}) and (\ref{momentum}), the Lorentz  algebra generator  takes the following form:
\begin{eqnarray}
    M^{\mu \nu}= \left(1- \alpha p_0\right) M_{0}^{\mu \nu}.
\end{eqnarray}
Consequently, the Poincaré algebra is expressed as:
\begin{eqnarray}
    &&[p_{\mu},p_{\nu}]=0\\
    &&\left[p^{\mu}, M^{\nu \rho}\right]= \left(1- \alpha p_0\right)~\left[p_0^{\mu}, M_0^{\nu \rho}\right]\\
    &&\left[M^{\mu \nu}, M^{\rho \sigma}   \right]=  \left(1- \alpha p_0\right)^2~\left[M_0^{\mu \nu}, M_0^{\rho \sigma} \right]\\
    &&\left[x^{\mu},M^{\nu \rho}\right]= \left(1- \alpha p_0\right) ~ [x_0^{\mu},M_0^{\nu \rho}]-  i \hbar \alpha \left(\frac{p_0^{\mu}}{p_0}+ 2 \alpha p_0^{\mu}\right) M_0^{\nu \rho}.\nonumber \\ \label{poincarex}
\end{eqnarray}
This modified Poincaré algebra is consistent with the principle of a minimum measurable length scale where the algebra approaches nullity at this minimal measurable scale ($p_{0}=1/\alpha$). Let us notice that using a definition of the electron degeneracy parameter $\Psi$ given as
\begin{equation}\label{degeneracy}
\psi:=\frac{k_{B} T}{E_{F}}=\frac{2 m_{e} k_{B} T}{\left(3 \pi^{2} \hbar^{3}\right)^{2 / 3}}\left[\frac{\mu_{e}}{\rho N_{A}}\right]^{2 / 3},
\end{equation}
where $k_B$ is the Boltzmann constant, $T$ temperature, and $E_F\approx p_0/2m_e$ Fermi energy for the non-relativistic electrons, one obtains that 
\begin{equation}
    \alpha_\text{sing} \sim E_F^{-1} \sim \rho^{-2/3} \sim -\mathcal R^{-2/3}.
\end{equation}
In physical systems, passing the singular value of the parameter provides a non-physical behavior (such as negative pressure) as discussed in \cite{Pachol:2023tqa,Wojnar:2023bvv,Lope-Oter:2023urz,Wojnar:2024xdy} and usually it serves as a lower bound for the physically admittable value of the parameter. That is, it can be used to constrain a theory. For example, tabletop experiments related to the behavior of the liquid helium provide the bound as 
$-0.25\times10^{23}\lesssim\alpha\lesssim 0.25\times10^{23}{\text{ s}}/{\text{kg m}}$ \cite{Wojnar:2024xdy} while the Earth's seismic data $-1.5\times10^{22}\lesssim\alpha\lesssim 0.75\times 10^{22}{\text{ s}}/{\text{kg m}}$ up to $2\sigma$ \cite{Wojnar:2023bvv,Kozak:2023ruu}. On the other hand, there are well-known situations in which one deals with the negative effective pressure, such as dark energy in cosmology (see e.g. \cite{copeland2006dynamics}) or negative pressure in trees (see e.g. \cite{mcelrone2013water,wheeler2008transpiration}). Due to that fact, stronger constraints resulting from the idealized models, like for example Bose-Einstein condensate \cite{Wojnar:2024xdy} or polytropic equation of state for stellar modeling \cite{barausse2008no,barausse2008curvature,Pachol:2023tqa,Wojnar:2023bvv}, should not be taken lightly and be a base to rule out given theories, as many important processes are not taken into account in such modeling.\\
\noindent
An exception to algebraic nullity manifests at the minimal length scale  ($p_{0}=1/\alpha$), as shown in Eq. (\ref{poincarex}), implying the existence of a non-null algebraic framework at this scale:
\begin{eqnarray}
 \left[x^{\mu},M^{\nu \rho}\right]= - 3 i \hbar \alpha^2 p_0^{\mu} M_0^{\nu \rho},
\end{eqnarray}
which can be contracted by Christoffel connection to give approximately:
\begin{equation}
    \Gamma_{\mu \nu}^{\rho}  \left[x^{\mu},M^{\nu}_{\rho}\right]  \sim - i \alpha  \hbar~.  \label{fixedgeometry}
\end{equation}
 This unique algebraic behavior in Eq. (\ref{fixedgeometry}) could be indicative of a background-independent algebra characteristic of quantum gravity theory \cite{Witten:2023xze}. Furthermore, by applying Eqs. (\ref{position}) and (\ref{momentum}), a modified dispersion relation is derived from the linear GUP, expressed as:
\begin{equation}
p^\mu p_\mu= p^2= p_0^2- 2 \alpha p_0^3 = m^2 c^2,
\end{equation}
where $m$ represents the mass of the particle.  This dispersion relation yields real momentum solutions up to the first order in $\alpha$, given by:
\begin{eqnarray}
p_{01} &= \frac{1}{2 \alpha} - 2~ \alpha~m^2 ~c^2,\label{solution1} \\
p_{02} &= m c + \alpha~m^2~c^2,\label{solution2}  \\
p_{03} &= -m c + \alpha~m^2~c^2. \label{solution3}
\end{eqnarray}
These solutions are valid under the condition that $27 ~m^2 c^2 ~\alpha^2 \geq 1$. This bound may point to sub-Planckian physics \cite{Carr:2015nqa,Ling:2021olm} if we set $m$ in the order of Planck mass $M_{Pl}$. In the absence of this condition, the solutions become complex, featuring imaginary components of the momentum. We notice here there exist three real solutions. One of these solutions presented by Eq. (\ref{solution1}) is non-perturbative and remains unaffected by the particle's mass. This particular solution is vital in the discretization of spacetime as evident in \cite{Ali:2009zq, Ali:2011fa, Das:2010zf,Das:2020ujn}.
The quantum gravity corrections highlighted in the solutions provided by Eq. (\ref{solution2}) and Eq. (\ref{solution3}) suggest possible impacts on a broader spectrum of physical systems. Specifically, these corrections could influence studies in condensed matter, high energy physics, and astrophysics, and extend to areas such as cosmology, quantum computing, and particle physics.

\section{ Covariant GUP and Covariant Entropy bound}\label{sec:GUP-Bekenstein}

\begin{figure}[ht!]
    \includegraphics[scale= 0.2]{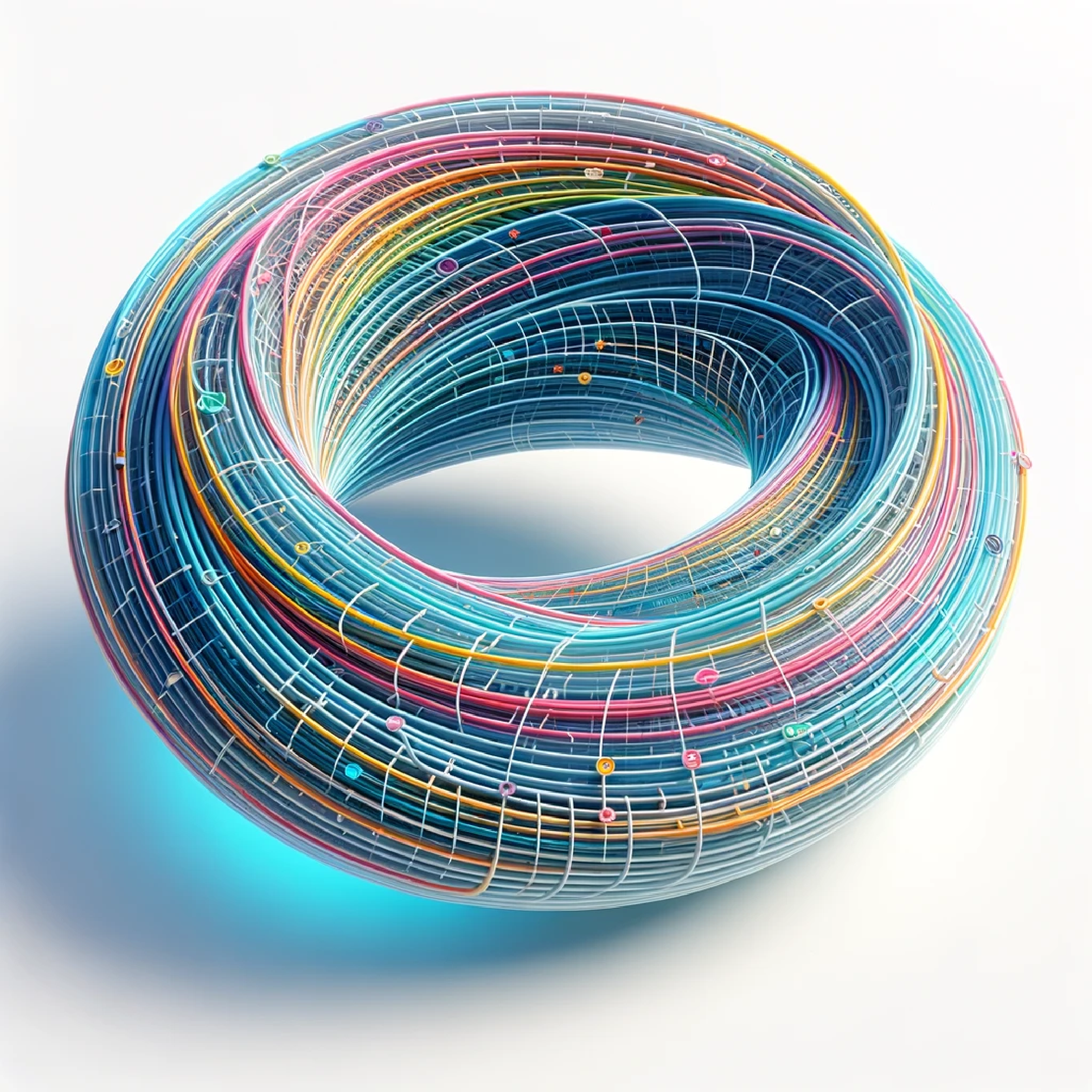}
    \caption{A 3-torus representing the Bekenstein bound on the physical system and consistent with GUP/minimal length models with $r_s$ as inner radius and $R$ as outer radius }
    \label{fig:torus}
\end{figure}

\noindent
Recent research has revealed a possible link between the Bekenstein bound inequality \cite{Bekenstein:1980jp}, which determines the maximal amount of information needed to precisely characterize a physical system upto the quantum level, and the uncertainty inequality \cite{Buoninfante:2020guu}. This study, however, simplified the entropy concept to unity to make it consistent with the uncertainty inequality. When retaining entropy as $S$, a likely equivalence between the Bekenstein bound inequality and the GUP is revealed, as illustrated in \cite{Ali:2022ckm, Ali:2022jna}. The entropic origin of GUP was also investigated in \cite{CaboBizet:2022hpz}. This hypothesized equivalence has been supported by explaining the radii of hydrogen atoms and nuclei, the cosmological constant's value as well as resolving the EPR paradox \cite{Ali:2022jna,AEGMV,Ali:2022ulp}. This exploration prompts further studies into the covariant form of the Bekenstein bound and its possible equivalency with the Covariant GUP we are investigating in this paper. In recent investigations \cite{Ali:2022ckm}, further illumination on this equivalence was achieved by noting distinct patterns in the Bekenstein bound value, denoted as \(MRc\), in relation to the Planck constant. For particles, \(MRc\) consistently aligned around the Planck constant, whereas for atoms or nuclei, it showed fluctuations within several orders of magnitude of the Planck constant. Here, \(M\) signifies the mass of the entity in question, and \(R\) is the radius of the smallest encompassing sphere. This led to the formulation of the equation:
\begin{eqnarray}
MRc = \alpha^{\prime} \hbar~. \label{universal}
\end{eqnarray}
The coefficient \(\alpha^{\prime}\) is derived from empirical measurements of particle mass and charge radius. Noteworthy examples include pions (\(MRc = 0.449 \hbar\)) and kaons (\(MRc = 1.458 \hbar\)) \cite{Ali:2022ckm}. The observed fluctuations in the vicinity of the Planck constant have led to the postulation of a correlation between the Bekenstein bound and the GUP. The GUP also suggests an effective variation of the Planck constant, as reflected in the modified commutation relations between position and momentum. The Bekenstein bound on the entropy can be formulated as:
\begin{eqnarray}
S \leq \frac{2 \pi~k_B R~ E}{\hbar~ c} ~.\label{Bekenstein}
\end{eqnarray}
If we assume that $E= M~c^2$, where $M$ is the relativistic mass, then the Bekenstein bound can be formulated as
\begin{eqnarray}
S \leq 2 \pi~k_B \frac{ M~ R~c}{\hbar} ~.\label{Bekenstein}
\end{eqnarray}
On other side, it was considered in \cite{Ali:2022ckm,Ali:2022jna} that,
\begin{eqnarray}
&& M~R~c= \alpha^{\prime} \hbar \label{shape} \implies \\ 
&& r_s~R = 2 \alpha^{\prime} \ell_{Pl}^2~.\label{shape1}
\end{eqnarray}
In this context, \(r_s= 2 M G/c^2\) represents the Schwarzschild radius associated with the physical object that has a relativistic mass $M$. The factor \(\alpha^{\prime}\), when comparing with Eq. (\ref{Bekenstein}), is reflective of Bekenstein entropy bound, and brings us to:
\begin{equation}
  \frac{S}{2\pi k_B} \leq \alpha^{\prime}  \label{S-alpha}   
\end{equation}
Incorporating  Eq. (\ref{shape1}) into Eq. (\ref{S-alpha}), we arrive at the following inequality:
\begin{equation}
   \frac{S}{\pi k_B} \ell_{Pl}^2 \leq r_s~ R~.
\end{equation}
Which can be formulated as follows:
\begin{eqnarray}
   \frac{S}{ \pi k_B} \leq \frac{A_{\text{3-Torus}}}{4 \pi^2 \ell_{Pl}^2}~, \label{Trous}
\end{eqnarray}
where \(A_{\text{3-Torus}} = 4 \pi^2 r_s~ R\). The area bound is interpreted via a 3-Torus, characterized by an inner radius \(r_s\) and an outer radius \(R\).
This formula is different from Bousso's bound that defines a covariant entropy bound \cite{Bousso:1999xy} that is given by:
\begin{eqnarray}
   \frac{S}{ \pi k_B} \leq \frac{A_{\text{Sphere}}}{4 \pi^2 \ell_{Pl}^2}~, \label{Bousso}
\end{eqnarray}
where \( A_{\text{Sphere}} = 4 \pi R^2 \) denotes the area of the smallest sphere that encloses the system. In contrast to Bousso's approach, which uses the gravitational stability condition \( r_s < R \) and replaces \( r_s \) with \( R \) in the Bekenstein inequality \ref{Bekenstein} in order to get Eq. (\ref{Bousso}), our methodology differs. To achieve an accurate geometric description of the Bekenstein universal bound, it suffices to simply use \( r_s \) in its original form. In our analysis, we retain \( r_s \) as it is and derive the corresponding area as that of a torus. For this torus, the outer radius is defined by the smallest sphere that encloses the physical system, and the inner radius is determined by the Schwarzschild radius associated with the system's mass. This approach allows our formulation to be in harmony with the Bekenstein bound and provides a self-sufficient mathematical basis for representing the covariant form of entropy bound in terms of the area of a 3-torus. To say it another way, the covariant entropy bound, which we determined as a torus area, offers a precise geometric interpretation of the Bekenstein universal bound. Bousso's approach relies on the area of the smallest sphere encompassing a physical system, which does not provide an exact geometric description of the Bekenstein universal bound. In reality, Bousso's bound tends to be significantly larger than the Bekenstein universal bound, in particular for macroscopic systems. The entropy bound we introduced integrates the minimal length of measurement concept and potentially reflects geometric aspects of the GUP. Our approach, as outlined in \cite{Ali:2022jna} introduces a role of entropy in the uncertainty principle as:
\begin{equation}
    \Delta x \Delta p \geq \frac{S}{ \pi k_B}~\frac{\hbar}{2} ~.\label{QG1}
\end{equation}
The importance of Eq. (\ref{QG1}) was discussed and verified by one of the authors by explaining the Hydrogen atom/nucleus radii, the cosmological constant value, and the EPR paradox \cite{Ali:2022ckm,Ali:2022jna,Jusufi:2023dix,Ali:2022ulp}, and now we wish to expand on its conceptualization. This equation integrates fundamental elements of nature: the things we can know (position and momentum) and the things we cannot know (entropy). Over the last few decades, entropy has been increasingly recognized as a cornerstone in understanding gravity and the Einstein field equations. This perspective originates from thermodynamic approaches to GR, which interpret the Einstein field equations as an equilibrium equation of state, derived from thermodynamic principles \cite{jacobson1995, wald2001}. However, let us notice that modified gravity, which introduces additional curvature terms and/or extra degrees of freedom, requires a non-equilibrium treatment\footnote{Due to that fact, one can deal with an extra term related to the entropy developed internally in the system as a result of being out of equilibrium in the Clausius' relation \cite{eling2006nonequilibrium}. We will come back to that problem in near future.} since entropy acquires the curvature corrections \cite{eling2006nonequilibrium,faraoni2021new,giusti2022first}. 
Central to this derivation is the entropy-area law \cite{bekenstein1973, hawking1975}, linking the entropy of a black hole to its event horizon area, used alongside the Clausius relation \(\delta Q = T dS\) (in GR), where \(\delta Q\) represents heat flow, \(T\) the Unruh temperature, and \(dS\) the change in entropy. The Raychaudhuri equation \cite{raychaudhuri1955} also plays a pivotal role in deriving the Einstein field equations \cite{jacobson1995}. Additionally, Verlinde's work \cite{Verlinde:2010hp,Verlinde:2016toy} interprets gravity as an entropic force, leading to the derivation of Newton's law of gravitation and the Einstein field equations. It is then naturally thought that the inequality in Eq. (\ref{QG1}) represents an expression of the quantum theory of gravity.\\
\noindent
Based on Eqs. (\ref{Trous}) and (\ref{QG1}),
a foundational equation reads:
\begin{equation}
    \Delta x \Delta p = \frac{A_{3-\text{Torus}}}{4\pi^2 \ell_{Pl}^2} \frac{\hbar}{2}~.\label{QGdiscovery1}
\end{equation}
The Eq. (\ref{QGdiscovery1}) sheds light on the geometric basis for the uncertainty relationship between position \(\Delta x\) and momentum \(\Delta p\) in standard quantum mechanics, traditionally expressed as an inequality \(\Delta x \Delta p \geq \frac{\hbar}{2}\). This same equation (\ref{QGdiscovery1}) reveals that the area of a 3-Torus, \( A_{3-\text{Torus}} \), plays a crucial role in transforming the uncertainty principle from an inequality into a specific equation. By incorporating \( A_{3-\text{Torus}} \), the principle is reframed as a definitive equation, provided the condition \( \frac{A_{3-\text{Torus}}}{4\pi^2 \ell_{Pl}^2} \geq 1 \) is met. This condition is satisfied under the constraints of gravitational stability, where \( r_s \leq R \), and in alignment with the minimal measurable length, as indicated by \( \alpha_0 \ell_{Pl} = r_s \) \cite{Scardigli:1999jh}. Bounds larger than 1 on the parameter $\alpha_0$ can be found from studies of GUP with several physical systems as investigated in a range of studies  \cite{Ali:2011fa, Bosso:2018ckz,Pikovski:2011zk,Kumar:2017cka, Tino:2020dsl,Gao:2016fmk, Das:2011tq,Marin:2013pga,Bosso:2018ckz,Feng:2016tyt,Petruzziello:2020wkd, Kumar:2019bnd, Moradpour:2019wpj,Vagnozzi:2022moj, Iorio:2017vtw,Bawaj:2014cda,Das:2021nbq}. Using  the Cauchy-Schwarz inequality, the covariant linear GUP takes the following form:
\begin{equation}
  [x_{\mu}, p_{\nu}] = i \hbar \left[ g_{\mu \nu} - \alpha \left( p g_{\mu \nu} + \frac{p_{\mu} p_{\nu}}{p} \right) \right] = i \hbar~ g_{\mu \nu} \frac{S}{ \pi k_B}~. \label{QGdiscovery}
\end{equation}
The Eqs. (\ref{QGdiscovery1}) and (\ref{QGdiscovery})  provide a foundational basis for further exploration into the geometric interpretation of the linear GUP, which can be simply extended to various covariant minimal length theories \cite{Amati:1988tn,Scardigli:1999jh,snyder1947quantized,Brau:1999uv,Adler:2001vs,maggiore1993generalized,Capozziello:1999wx,Kempf:1994su}. A figure showing the 3-Torus is given in Fig. (\ref{fig:torus}). 
The torus is typically described using Riemannian geometry \cite{lee2000riemannian}. Its curved structure, which cannot be made flat without distortion, contrasts with the flat spacetime of Special Relativity where space is not curved by gravity or mass \cite{mtw1973gravitation}. 
 The linear GUP modeled through a 3-torus, embedded in Riemannian geometry, not only demonstrates compatibility with modified gravity theories but also illuminates a critical linkage in Riemannian geometry, as indicated by Eqs. (\ref{QGdiscovery1}) and (\ref{QGdiscovery}). This geometric interpretation facilitates a broader comprehension of quantum gravity's intricacies. The adoption of this model allows for an in-depth analysis of gravity's role within quantum systems, contributing to a more nuanced understanding across disciplines such as quantum field theory, cosmology, material science, and quantum information. Thus, employing a 3-torus to represent GUP offers a geometric approach for exploring quantum gravity's effects across diverse physical phenomena.

\section{Conclusions} \label{sec:conclusions}
\noindent
In conclusion, we have developed a covariant version of a linear GUP model for Palatini-like theories of gravity. This is grounded in the recently proposed correspondence between metric-affine gravity and linear GUP at the phase space level. Consequently, we were able to derive the corresponding modified Poincaré algebra. This algebra exhibits distinct behavior, especially in the vicinity of nullity at the minimal length scale indicated by linear GUP, except for the commutation relation between position and Lorentz generator. This non-vanishing commutation relation may point to a fixed background algebra in quantum gravity. A key element of our research is the adoption of 3-torus geometry for a covariant geometric visualization of GUP, which precisely describes geometrically the Bekenstein universal bound. Our approach retains the Schwarzschild radius (\( r_s \)) in its original form, diverging from the norm of substituting it with the radius (\( R \)) of the smallest sphere enclosing the system in the Bekenstein bound inequality. This leads to a re-envisioned covariant entropy bound, represented by the area of a torus with the inner radius determined by the Schwarzschild radius and the outer radius matching the smallest sphere enclosing the physical system. In our modest attempt, we propose a method that envisions the covariant entropy bound as the area of a torus, aiming to provide a more accurate geometric description when compared to Bousso's covariant entropy bound. While Bousso's bound, which is based on the area of the smallest sphere enclosing a physical system, offers valuable insights, our approach seeks to refine the geometric representation of the Bekenstein universal bound. We acknowledge that in certain cases, particularly with macroscopic systems, Bousso's bound becomes larger than the Bekenstein universal bound, and our method is an effort to address this disparity. Additionally, in a tentative step forward, we suggest an equation that combines the linear covariant GUP with the 3-torus covariant entropy bound. This may open a new avenue for further exploration of quantum gravity. { Moreover, the mentioned correspondence between modified gravity and linear GUP enables us to leverage tools developed by both communities for studying physical processes across large and small scales. For example, we can test the GUP using seismic data \cite{Kozak:2023vlj,Wojnar:2024xdy}, a method designed to constrain modified gravity theories. Conversely, the quantum-thermodynamic description of the non-relativistic limit of modified gravity allows us to test gravity models in tabletop experiments \cite{Petruzziello:2021vyf,Wojnar:2024xdy}. Connecting (modified) gravity with mathematical structures describing the quantum properties of a given system, as provided in this paper, represents a fundamental step towards advancing us closer to formulating a quantum theory of gravity. Investigating these potential effects across such a diverse range of physical systems presents an exciting direction for future research. We hope to report on these in the future.

\section*{Acknowledgments}
 AW acknowledges financial support from MICINN (Spain) {\it Ayuda Juan de la Cierva - incorporaci\'on} 2020 No. IJC2020-044751-I.\\
 In memory of Grzesiek Burda, you have become one of those stars we often spoke about during our teenage years.

\bibliographystyle{apsrev4-1}
\bibliography{ref.bib}{}

\end{document}